\documentclass[aps,twocolumn,superscriptaddress]{revtex4}

\usepackage{amsmath}
\usepackage{amssymb}
\usepackage{braket}
\usepackage{graphicx}
\usepackage{ulem}
\usepackage{xcolor}
\usepackage{xspace}

\newcommand{\rmi}{\mathrm{i}}
\newcommand{\rmd}{\mathrm{d}}
\newcommand{\Rmat}{\ensuremath{R_\omega}}
\definecolor{darkgreen}{rgb}{0,.7,0}

\begin{document}
\title{Strong-field photoionization by circularly polarized light}

\author{Jonathan Dubois}
\affiliation{Sorbonne Universit\'{e}, CNRS, Laboratoire de Chimie Physique – Mati\`{e}re et Rayonnement, LCPMR, 75005 Paris, France}
\affiliation{Max Planck Institute for the Physics of Complex Systems, N\"{o}thnitzer Stra\ss{e} 38, 01187 Dresden, Germany}
\author{Camille L\'{e}v\^{e}que}
\affiliation{Sorbonne Universit\'{e}, CNRS, Laboratoire de Chimie Physique – Mati\`{e}re et Rayonnement, LCPMR, 75005 Paris, France}
\author{J\'{e}r\'{e}mie Caillat}
\affiliation{Sorbonne Universit\'{e}, CNRS, Laboratoire de Chimie Physique – Mati\`{e}re et Rayonnement, LCPMR, 75005 Paris, France}
\author{Richard Ta\"{i}eb}
\affiliation{Sorbonne Universit\'{e}, CNRS, Laboratoire de Chimie Physique – Mati\`{e}re et Rayonnement, LCPMR, 75005 Paris, France}
\author{Ulf Saalmann}
\affiliation{Max Planck Institute for the Physics of Complex Systems, N\"{o}thnitzer Stra\ss{e} 38, 01187 Dresden, Germany}
\author{Jan-Michael Rost}
\affiliation{Max Planck Institute for the Physics of Complex Systems, N\"{o}thnitzer Stra\ss{e} 38, 01187 Dresden, Germany}

\begin{abstract}
We demonstrate that strong-field ionization of atoms driven by circularly polarized light becomes an adiabatic process when described in the frame rotating with the laser field. As a direct consequence, a conservation law emerges: in the rotating frame the  energy of the tunneling electron is conserved for rotationally invariant potentials. 
This conservation law, arising from a classical picture, is retrieved through a proper classical-quantum correspondence  when considering the full quantum system, beyond the Strong Field Approximation.
\end{abstract}

\maketitle

\paragraph*{Introduction}
Tunnel ionization is a fundamental quantum process which plays a key role in probing techniques for measuring the real-time motion of electrons inside atoms and molecules~\cite{Meckel2008, Corkum2011, Kling2013, Peng2015, Lin2018}. 
In order to probe experimentally this dynamics on an attosecond timescale, infrared and near-infrared laser pulses are most commonly employed. 
When the laser intensity is strong enough~\cite{Keldysh1965, Smirnov1966, Reiss1980}, a valence electron of an atom or molecule can tunnel ionize through the potential barrier induced by the non-perturbative field. The subsequent photoelectron dynamics, governed by field-driven rescattering~\cite{Corkum1993}, lead to highly non-linear phenomena such as above-threshold ionization~\cite{Paulus1994} or high-harmonic generation~\cite{Huillier1993} that have been subsequently developed to design ``self-probing'' spectroscopies with unprecedented time and space resolutions~\cite{Lin2010, Haessler2011}. Controlling the conditions under which tunnel ionization occurs, predicting the phase-space configuration of the photoelectron wavepacket after tunneling, and modelling the tunneling rates, are essential theoretical steps for interpreting and decoding the experimental measurements which allow  one to retrieve attosecond-resolved information on electron dynamics.

The essence of tunnel ionization is efficiently captured by adiabatic, quasi-static, theories such as ADK~\cite{Keldysh1965, Ammosov1986, Arissian2010}.
However, in the infrared regime, the characteristic tunneling time of the electron and the laser period are on the same order of magnitude, such that energy of the electron under the barrier are significantly affected by nonadiabatic sub-cycle couplings~\cite{PerelomovI1966, PerelomovII1967, PerelomovIII1967, Boge2013}. 
As a consequence, the energy of the electron during tunnel ionization is strongly influenced by the oscillations of the laser field, and the tunneling electron gains energy on the order of an electron-volt~\cite{Barth2011, Klaiber2015, Ni2018}. These energy changes of the electron upon tunneling are called nonadiabatic effects~\cite{Boge2013} and the energy of the electron right after tunneling is hard to assess.
In atoms, this is commonly achieved by neglecting the interaction between the electron and the ion~\cite{PerelomovI1966, PerelomovII1967, PerelomovIII1967, Barth2011} in the framework of the so-called  strong-field approximation~\cite{Lewenstein1994, Amini2019} (SFA). 
The SFA not only provides analytic formulas for ionization phenomena, it also unravels the classical behavior of the electron subjected to strong-laser fields. It is thus an essential ingredient for the design and  interpretation of time-resolved experiments using intense laser fields.

While the essential of strong-field physics can be addressed by considering linearly polarized pulses, circularly polarized (CP) fields valuably offer additional experimental ways of probing ultrafast dynamics in atoms and molecules, as demonstrated with the ``Attoclock'' setup~\cite{Eckle2008, Eckart2018}. 
There, information on the target and on the tunneling process is directly encoded in the photoelectron momentum distributions~\cite{Sainadh2019} since driving tunnel-ionized electrons with CP fields dramatically reduces recollision.
However, the semiclassical treatment of quantum strong-field tunneling has raised several debates on the time spent by the electron under the potential barrier~\cite{Torlina2015, Ni2016, Sainadh2019, Hofmann2021}. 
It is shown that the ion-electron interaction plays a crucial role and cannot be overlooked.
While intense CP light presents a unique avenue for probing chirality of molecules, a major obstacle remains to predict and control the phase-space configuration of the electron by fully taking into account the ion-electron interaction and nonadiabatic effects during tunnel ionization.

In this letter, we show that tunnel ionization of electrons by CP fields obeys a classical conservation law. By choosing a proper reference frame and by fully taking into account the ion-electron interaction, the nonadiabatic effects occuring on short timescales in the laboratory frame (LF) are transformed into adiabatic ones. Making use of this conservation law unravels the intrinsic link between the angular momentum of the electron and its energy variations in CP fields. We also show that this conservation law is, in fact, present in the SFA, and notably supports the classical picture of the process. Atomic units are used unless stated otherwise.

\begin{figure}
	\centering
    \includegraphics[width=.5\textwidth]{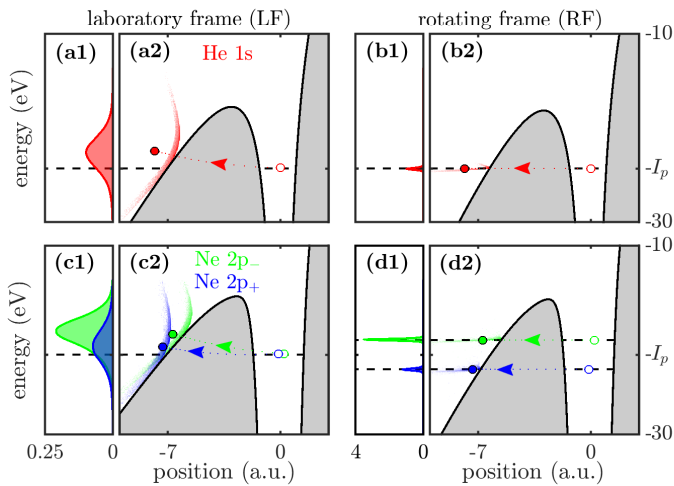}
	\caption{Configuration of the electron after tunnel ionization obtained by the backpropagation method~\cite{Ni2016, Ni2018}. (a) and (c) are the distributions in the LF. (b) and (d) distributions in the RF. Left panels are the distributions of the energy after tunneling (normalized with respect to their ionization probability), right panels are the distributions in energy and in position along the field direction ($\mathbf{F}(t)$ or $\widetilde{\mathbf{F}}(t)$) after tunneling. The grey regions indicate the classically forbidden region of the electron in the LF and in the RF at the peak amplitude of the laser field (i.e., at time $t{=}0$). The upper panels are for $\mathrm{He}$ for the initial state $1{\rm s}$ (red) for $I {=} 8 {\times} 10^{14} \; \mathrm{W}\ \mathrm{cm}^{-2}$, and the lower panels are for $\mathrm{Ne}$ for the initial state $2{\rm p}_-$ (green) and $2{\rm p}_+$ (blue) for $I {=} 6 {\times} 10^{14} \; \mathrm{W}\ \mathrm{cm}^{-2}$. The dotted curves are complex trajectories of tunneling electron with initial (open circles) and final conditions (solid circles).}
	\label{fig:LF_RF}
\end{figure}

\paragraph*{Laboratory frame}
We consider a single-active electron in an atom interacting with a classical electric field in the dipole approximation within the length gauge. The Hamiltonian governing the dynamics is
\begin{equation}
\label{eq:Hamiltonian_LF}
{\rm H}(t) = \dfrac{\mathbf{p}^2}{2} + V(\mathbf{r}) + \mathbf{r} \cdot \mathbf{F}(t) ,
\end{equation}
where $\mathbf{r}$ is the electron position, $V(\mathbf{r})$ is the ion-electron energy potential and $\mathbf{p}{=}{-}\rmi \boldsymbol{\nabla}$ is the momentum operator. 
The Hamiltonian~\eqref{eq:Hamiltonian_LF} is here expressed in the LF, i.e. the frame in which the electrons are detected in experiments. 
We consider ion-electron energy potentials which are invariant under rotations, which typically corresponds to atomic potentials~\cite{Tong2005} (and to some extent to molecules modeled by a continuous potential such as benzene~\cite{Xie2008} and buckminsterfullerene~\cite{Hertel2005}). We used soft-Coulomb potentials~\cite{Javanainen1988} with parameters adapted to model He~\cite{Mauger2014} and Ne~\cite{Barth2014} in the single active electron approximation  (see Supplemental Material~\cite{SuppMat}). To address the specificities of CP driven dynamics, we performed simulations for He  initially in its 1s ($m{=}0$)  state (with ionization potential $I_p {=} 24.3 \, {\rm eV}$), or Ne initially in each of its ${\rm 2{\rm p}_{\pm}}$ ($m{=}{\pm}1$) states (with ionization potential $I_p {=} 21.6 \, {\rm eV}$). 
The time-dependent CP laser electric field is defined as $\mathbf{F}(t) {=} {-} \partial_t \mathbf{A}(t)$ where $\mathbf{A}(t)$ is the associated vector potential 
\begin{equation}
\label{eq:vector_potential}
\mathbf{A}(t) = \dfrac{F}{\omega} f(t) \Big( \mathbf{e}_x \cos (\omega t) + \mathbf{e} _y \sin (\omega t)\Big). 
\end{equation}
Considering $\mathbf{e}_z$ (orthogonal to the polarization plane) as the quantification axis, a $m{=}{+}1$ (${-}1$) electron is therefore co-rotating (counter-rotating) with the laser field.
The amplitude of the laser is $F$, its intensity is $I {=} 2 F^2$ and its frequency is $\omega$. Here, we consider an IR of 800 nm and a 2-cycle $\sin^4$ envelop given as $f(t) {=} \cos (\pi t / \tau)^4$ for $|t| {\leq} \tau/2$ and zero otherwise, with $\tau {=} 4{\times}2\pi{/}\omega$. 

The $(\mathbf{r},\mathbf{p})$ phase-space distributions, right after tunneling and fully taking into account the ion-electron interaction, are obtained using the backpropagation method~\cite{Ni2016, Ni2018}. For this, the wavefunction $\psi (\mathbf{r},t)$, starting from the initial state $\psi_0(\mathbf{r})$, is propagated forward quantum mechanically using the time-dependent Schr\"{o}dinger equation (TDSE) $\rmi \partial_t \psi {=} {\rm H}(t) \psi$,  until one cycle after the end of the laser pulse, i.e. $T{=}3{\times}2\pi{/}\omega$. From $\psi (\mathbf{r},T)$, we can extract the  classical phase-space distribution at $T$, which is then propagated backward, using Hamilton's equations until we match the tunneling condition, corresponding to the vanishing of the longitudinal momentum ~\cite{Ni2016, Ni2018}. These equations are defined from the classical analog of ${\rm H}$ (Eq.~\eqref{eq:Hamiltonian_LF}), hereafter referred to as ${\cal H}(\mathbf{r},\mathbf{p},t)$ (all through the text, calligraphic letters stand for classical analogs).  

Figure~\ref{fig:LF_RF} displays the reconstructed energy distributions in the He case [frames (a), red] and in the Ne cases [frames (c), green ($m{=}{-}1$) and blue ($m{=}{+}1$)].
\begin{table}
    \centering
    \begin{tabular}{r | c c c}
         Atom\hspace{1mm} & \hspace{1mm}  He ${\rm 1s}$ \hspace{1mm}  & \hspace{1mm} Ne ${\rm 2p_-}$ \hspace{1mm} & \hspace{1mm} Ne ${\rm 2p_+}$  \\
         \hline
         $e$ (eV)  & 1.642 & 2.362 & 0.958  \\
         $\Delta e$ (eV)  & 2.064 & 1.839 & 2.070 \\ 
         \hline
         $\widetilde{e}$ (eV)  & $-0.032$ & $0.048$ & $-0.013$ \\
         $\Delta \widetilde{e}$ (eV)  & ${\color{white}+}0.113$ & $0.102$ & ${\color{white}+}0.120$ \\ 
         \hline
         $\theta$ (deg.)  & 0.452 & 1.331 & 0.205 \\
         $\Delta\theta$ (deg.)  & 6.728 & 7.561  & 6.632 \\
    \end{tabular}
    \caption{Statistical quantities characterizing the energy and angular distributions of the electron after tunneling in the laboratory frame and in the rotating frame obtained by the semi-classical backpropagation method (see text). In the former, the distribution of $E {+} I_p$ is fitted by a gaussian with mean value $e$ and standard deviation $\Delta e$. In the latter, the distribution of $\widetilde{E} {+} (I_p {+} m \omega)$ is fitted by a gaussian with mean value $\widetilde{e}$ and standard deviation $\Delta\widetilde{e}$. The distribution of $\sphericalangle({-}\mathbf{r},\mathbf{F}(t))$ is fitted by a gaussian with mean value $\theta$ and standard deviation $\Delta \theta$.}
    \label{tab:fit_error}
\end{table}
Four main features emerge from this figure. 
First, we see in frame (c1) that the ionization probability is larger for an initially counter-rotating electron than for a co-rotating one, in agreement with~\cite{Barth2011, Eckart2018}.
Second, frames (a1) and (c1) show that, starting from a delta-function at ${-}I_p$, the energy distribution shifts towards higher energies (by around 1 eV) and gets a width of about 2 eV during the tunneling process. This is a clear signature of non-adiabatic effects and was also observed in~\cite{Ni2018}. 
Third, frame (c1) also shows that, for oriented initial states, the photoelectron peak shifts towards higher energies roughly twice more, and with a smaller width, for $m{=}{-}1$ than for $m{=}{+}1$. 
These three first features are supported by the quantitative data reported in the first two lines of table~\ref{tab:fit_error}.
Finally, in frames (a2) and (c2) we observe that the obtained comma-shaped electron distributions in the energy-position plane lie close to the potential barrier at the peak amplitude of the laser field (the classically forbidden regions are indicated by grey areas), regardless of its initial energy and its magnetic quantum number. 
These large energy variations during tunneling can be assessed either using  SFA~\cite{Barth2011}, and thus neglecting the ion-electron interaction, or  going into the rotating frame  where we can interpret them on subcycle timescales. 

\paragraph*{Rotating frame perspective}
The rotating frame (RF) has already been used either in the classical context for strong-field physics in~\cite{Mauger2010, Kamor2013}, or in the quantum context for high-harmonic generation by bicirculary polarized pulses~\cite{Reich2016} and for ionization by microwaves~\cite{Buchleitner2002}. Switching to the frame which rotates with the laser field is formally achieved by means of the time-dependent matrix $\Rmat(t)$ associated with the rotation of angle $\omega t$ around $\mathbf{e}_z$. In the RF associated with the CP field, the vector potential is $\widetilde{\mathbf{A}}(t) {=} \Rmat(t)\mathbf{A}(t) {=} (F/\omega) f(t) \mathbf{e}_x$ and the laser electric field is $\widetilde{\mathbf{F}}(t) {=} \Rmat(t) \mathbf{F}(t) {=} {-} (F/\omega)(\dot{f}(t)\mathbf{e}_x + f(t) \omega \mathbf{e}_y)$.  The fast carrier oscillations at frequency $\omega$ present in the LF disappear in the RF. 

Wavefunctions $\widetilde{\psi}$ in the RF frame are related to ${\psi}$ in the LF frame by the unitary transformation
\begin{equation}
\label{eq:transformation_LF_RF}
    \widetilde{\psi} (\mathbf{r} , t) = \exp \left( \rmi \, \omega t \,  {\rm L}_z \right) \psi (\mathbf{r},t) \equiv \psi( \Rmat^{-1}(t) \mathbf{r},t) ,
\end{equation}
where ${\rm L}_z {=} \mathbf{r} {\times} \mathbf{p} {\cdot} \mathbf{e}_z$ is the angular momentum normal to the polarization plane and $\widetilde{\psi}$ is the wavefunction of the electron in the RF.
In the RF, the TDSE becomes $\rmi \partial_t \widetilde{\psi} {=} \widetilde{\rm H}(t)\widetilde{\psi}$ with Hamiltonian 
\begin{equation}
    \label{eq:Hamiltonian_RF}
    \widetilde{\rm H} (t) = \dfrac{\mathbf{p}^2}{2} + V(\mathbf{r}) - \omega {\rm L}_z + \mathbf{r}\cdot \widetilde{\mathbf{F}} (t) ,
\end{equation}
where the Coriolis term $\omega {\rm L}_z$ results from the time-dependent rotation~\eqref{eq:transformation_LF_RF} from the LF to the RF. 
Due to the rotational invariance of Hamiltonian~\eqref{eq:Hamiltonian_RF} in absence of laser field, it shares the same field-free eigenstates as Hamiltonian ${\rm H}$~\eqref{eq:Hamiltonian_LF} but shifted in energy due to the Coriolis term. Thus, the ionization potential in the RF is $\widetilde{I}_p {=} I_p {+} m \omega$ with $\widetilde{\rm H}({-}\infty)\psi_0 {=}{-}\widetilde{I}_p \psi_0$.
The classical Hamiltonian in the RF is denoted $\widetilde{\cal H}(\widetilde{\mathbf{r}},\widetilde{\mathbf{p}},t)$ where $\widetilde{\mathbf{r}}$ and $\widetilde{\mathbf{p}}$ are the phase-space variables of the electron in the RF.
Hamiltonian $\widetilde{\cal H}$ is obtained either by using the classical analogy of~\eqref{eq:Hamiltonian_RF}, or equivalently by performing the time-dependent canonical transformation $\widetilde{\mathbf{r}}{=} \Rmat(t) \mathbf{r}$ and $\widetilde{\mathbf{p}}{=} \Rmat(t) \mathbf{p}$ from ${\cal H}$. 

\begin{figure}
	\centering
	\includegraphics[width=.5\textwidth]{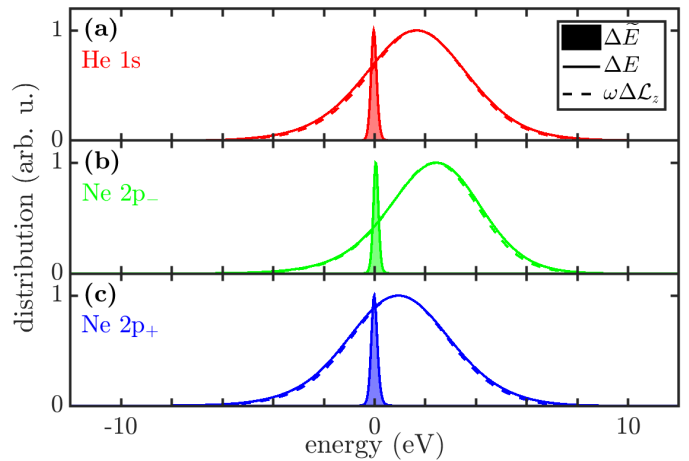}
	\caption{Distributions of the energy variation in the rotating frame $\Delta \widetilde{E} {=} \widetilde{E}{+} (I_p {+} m\omega)$ (filled lines), in the laboratory frame $\Delta E {=} E{+} I_p$ (solid lines) and distribution of the inertial energy variation $\omega\Delta {\cal L}_z {=} \omega({\cal L}_z {-} m)$ (dashed lines), corresponding to the distributions of Fig.~\ref{fig:LF_RF}(.1) shifted by their initial energy and normalized to unity. All the data were obtained by the backpropagation method and normalized to unity. The gaussian fit parameters of $\Delta E$ and $\Delta \widetilde{E}$ are given in table~\ref{tab:fit_error}.}
	\label{fig:energy_variations}
\end{figure}

In the RF, the time-dependence of the electric field is reduced to its envelop $f(t)$, which (i) varies on timescales longer than a laser cycle and (ii) can play the role of an {\it adiabatic parameter}. Since tunnel ionization occurs on timescales shorter than the laser cycle, the classical picture predicts that the energy of the electron is approximately conserved during tunnelling.
Hence, we expect in the RF after tunneling the energy
\begin{equation}
    \label{eq:adiabatic_invariant}
    \widetilde{E} \approx - \big( I_p + m \omega \big) .
\end{equation}
This is indeed confirmed by the results of the backpropagation method displayed in Fig.~\ref{fig:energy_variations} where we show the distribution of energy gained by the electron {\it during} tunnel ionization in the RF, i.e. $\Delta \widetilde{E}{=} \widetilde{E}{+} (I_p{+}m\omega)$. In table~\ref{tab:fit_error}, we report for all initial states a shift of the photoelectron peak position in the RF ($\widetilde{e}$) of a few tens of meV, i.e. two orders of magnitude lower than in the LF ($e$). The peak width in the RF ($\Delta \widetilde{e}$) is around 100 meV, i.e. one order of magnitude lower than in the LF ($\Delta e$).
We have checked that the conservation law given in Eq.~\ref{eq:adiabatic_invariant}, within the tunneling regime, is robust with respect to field intensities and frequencies.
This confirms that tunnel ionization in the RF occurs adiabatically on the energy isosurface $\widetilde{\mathcal{H}}(\widetilde{\mathbf{r}},\widetilde{\mathbf{p}},t){=}\widetilde{E}$ in phase space, where $\widetilde{E}$ results from the $m$- and $\omega$-dependent electron-ion coupling and laser interactions, see Eq.~\ref{eq:adiabatic_invariant}.  
Note that this is a clear confirmation from TDSE calculations of the extension to the adiabatic regime of the ADK assumption that tunnel ionization occurs on a constant energy surface.

In the RF we can define an effective potential~\cite{SuppMat} as
\begin{equation}
\label{eq:Veff}
    \widetilde{V}_{\rm eff} (\mathbf{r},t) = V(\mathbf{r}) - \dfrac{\omega^2}{2} \, ( \mathbf{e}_z \times  \mathbf{r})^2 + \mathbf{r}\cdot \widetilde{\mathbf{F}}(t) .
\end{equation}
The classically forbidden regions where tunnelling takes place, at the peak of the pulse envelop (grey areas in Figs.~\ref{fig:LF_RF}b and~\ref{fig:LF_RF}d), are bounded by $\widetilde{V}_{\rm eff} (\mathbf{r},0)$. These regions dominate the tunneling dynamics.
As observed in Fig.~\ref{fig:LF_RF}d, 
the RF potential barrier is thinner and the energy gap between the top of the barrier and $\widetilde{I}_p$ is smaller for a ${\rm 2{\rm p}_{+}}$ electron (smaller $\widetilde{I}_p$) than for a ${\rm 2{\rm p}_{-}}$ one (larger $\widetilde{I}_p$). 
Since tunneling is strongly suppressed with increasing classically forbidden area, the ionization probability for a counter-rotating electron is larger than for a co-rotating one in strong CP fields in this typical regime, in agreement with the SFA~\cite{Barth2011}, numerical simulations~\cite{Barth2014} and experimental measurements ~\cite{Eckart2018}. 

\begin{figure}
	\centering
	\includegraphics[width=.5\textwidth]{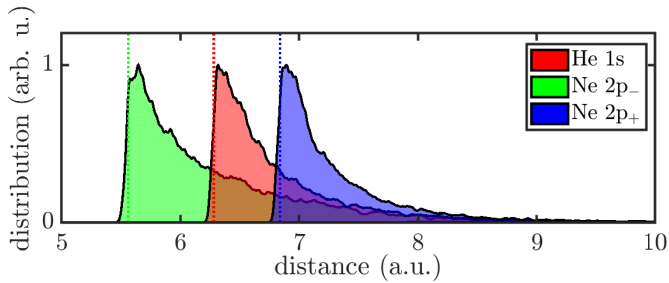}
	\caption{Distributions of the distance of birth of the electron from the origin obtained by the backpropagation method and normalized to unity. The vertical dotted lines are the position of the effective potential barrier $r_0$ computed from $\widetilde{V}_{\rm eff} (\mathbf{r},0) {=}{-} \widetilde{I}_p$ (see Eq.~\eqref{eq:Veff}) with tunneling exit at the peak amplitude of the laser field $\mathbf{r}{=}{-}r_0 \mathbf{F}(0) / |\mathbf{F}(0)|$.}
	\label{fig:distance}
\end{figure}

The last two lines of table~\ref{tab:fit_error} reveal that the electron ionizes mainly along the laser electric field direction, as predicted from tunneling theories~\cite{PerelomovI1966, PerelomovII1967, PerelomovIII1967, Goreslavski2004, Barth2011, Klaiber2015}, and that the position of the electron after tunneling is approximately given by $\widetilde{\mathbf{r}} {=} {-} r_0 \widetilde{\mathbf{F}}(t)/|\widetilde{\mathbf{F}}(t)|$. The radius $r_0$ can be determined using the conservation law in the RF (Eq.~\ref{eq:adiabatic_invariant}). It corresponds to the position of the outermost intersection between the effective potential $\widetilde{V}_{\rm eff} (\mathbf{r},t)$ and the initial energy ${-}\widetilde{I}_p$ , i.e. the solution of $\widetilde{V}_{\rm eff} ({-}r_0 \widetilde{\mathbf{F}}(t)/|\widetilde{\mathbf{F}}(t)|,t) {=} {-}\widetilde{I}_p$. 
In Fig.~\ref{fig:distance}, $r_0$ at the peak of the envelop is indicated by vertical dotted lines: it quantitatively matches the minimum exit distance after tunneling, given by the distributions plotted on the same figure.
Finally, in the RF, the momentum of the electron, when exiting the potential barrier, is $\widetilde{\mathbf{p}}{=} \omega \mathbf{e}_z {\times} \widetilde{\mathbf{r}}$ (zero kinetic energy, see also~\cite{Dubois2020}). It is therefore perpendicular to the electric field and non-zero, in agreement with tunneling theories~\cite{PerelomovI1966, PerelomovII1967, PerelomovIII1967, Barth2011}.

The knowledge gained from the RF perspective can now be used to deepen our understanding of both the phase-space configuration after tunneling in the LF, and the classical-quantum correspondence of tunnel ionization.
As at each time ${\cal H} (\mathbf{r},\mathbf{p},t) {=} \widetilde{\cal H} (\widetilde{\mathbf{r}},\widetilde{\mathbf{p}},t) {+} \omega {\cal L}_z$ and according to Eq.~\eqref{eq:adiabatic_invariant}, the angular momentum variations of the electron during tunnel ionization are directly converted into energy in the LF. Thus, we have 
\begin{equation}
    \label{eq:energy_conversion}
    \Delta E \approx \omega \, \Delta {\cal L}_z ,
\end{equation}
with $\Delta E{=} E {+}I_p$ and $\Delta {\cal L}_z{=} {\cal L}_z{-}m$ the energy and angular momentum variations of the electron before and after tunneling. 
Figure~\ref{fig:energy_variations} shows a comparison between $\Delta E$ and $\omega\Delta {\cal L}_z$, and the perfect agreement between the two curves confirms our findings resulting in the conversion law~\eqref{eq:energy_conversion}.

\paragraph*{Strong-field approximation}
Finally, we aim to validate the conservation law~\eqref{eq:adiabatic_invariant}, originating from a classical picture, with the commonly used SFA approach~\cite{Lewenstein1994, Milossevic2002, Amini2019}. 
We start in the LF from Hamiltonian~\eqref{eq:Hamiltonian_LF}. Within SFA , one uses an ansatz for the electronic wavefunction as the sum of the initial bound state $\psi_0$ and an ionized wavepacket $\varphi$, i.e. $\psi (\mathbf{r},t) {=} \exp (\rmi I_p t) \psi_0 (\mathbf{r}) {+} \varphi (\mathbf{r},t)$.
After substituting this approximation in the TDSE, one obtains 
\begin{subequations}
\label{eq:SFA}
\begin{equation}
\label{eq:TDSE_ionizing-wavepacket}
    \left( \rmi \,\partial_t - \dfrac{\mathbf{p}^2}{2} - \mathbf{r} \cdot \mathbf{F}(t) \right) \varphi (\mathbf{r},t) = s(\mathbf{r},t),
\end{equation}
where the ion-electron interaction on the ionizing wavepacket is neglected~\cite{Lewenstein1994} and 
\begin{equation}
    s(\mathbf{r},t) =  \big( \mathbf{r}\cdot \mathbf{F}(t) \big) \, \exp (\rmi I_p t ) \, \psi_0 (\mathbf{r}),
\end{equation}
\end{subequations}
is the electron source of the ionizing wavepacket which solely depends on the initial state~\footnote{The SFA equations~\eqref{eq:SFA} can be equivalently written in the length gauge or velocity gauge, in the position representation or momentum representation.}. 
In the Green function formalism, the dynamics of the ionizing wavepacket is then governed by
\begin{equation}
\label{eq:varphi_LF}
    \varphi (\mathbf{r},t) = \int_{-\infty}^t \rmd t^{\prime} \int \rmd \mathbf{r}^{\prime} \; G(\mathbf{r},t ; \mathbf{r}^{\prime},t^{\prime}) \, s(\mathbf{r}^{\prime},t^{\prime}) .
\end{equation}
The Green function $G(\mathbf{r},t ; \mathbf{r}^{\prime},t^{\prime})$ can be expressed exactly in terms of the classical action ${\cal S} (\mathbf{r},t ; \mathbf{r}^{\prime},t^{\prime})$ solution of the Hamilton-Jacobi equation since ${\cal H}$ is linear in position for $V{=}0$~\cite{Littlejohn1992, Littlejohn1986, SuppMat}.
In the RF, the dynamics of $\varphi$ is obtained by performing the transformation~\eqref{eq:transformation_LF_RF}
\begin{subequations}
\begin{equation}
    \label{eq:varphi_RF}
    \widetilde{\varphi} (\mathbf{r},t) = \int_{-\infty}^t \rmd t^{\prime} \int \rmd \mathbf{r}^{\prime} \; \widetilde{G}(\mathbf{r},t ; \mathbf{r}^{\prime},t^{\prime}) \, \widetilde{s} (\mathbf{r}^{\prime},t^{\prime}) ,
\end{equation}
with the source term $\widetilde{s} (\Rmat(t)\mathbf{r},t) {=} s \big( \mathbf{r},t \big)$. There, the initial state accumulates a time-dependent phase, i.e. $\psi_0(\Rmat^{-1}(t)\mathbf{r} ){=} \exp (\rmi m \omega t )\psi_0(\mathbf{r})$. Hence the source term becomes
\begin{equation}
    \widetilde{s} (\mathbf{r},t) =  \big( \mathbf{r} \cdot \widetilde{\mathbf{F}}(t) \big) \,  \exp \big( \rmi \widetilde{I}_p t\big) \,  \psi_0 (\mathbf{r}).
\end{equation}
\end{subequations}
The Green function becomes $\widetilde{G} (\Rmat(t) \mathbf{r},t;\Rmat(t^{\prime})\mathbf{r}^{\prime},t^{\prime}) {=} G\big(\mathbf{r} ,t ; \mathbf{r}^{\prime} ,t^{\prime}  \big)$, or equivalently the classical action becomes $\widetilde{\cal S} (\Rmat(t)\mathbf{r},t;\Rmat(t^{\prime})\mathbf{r}^{\prime},t^{\prime}) {=} {\cal S}\big(\mathbf{r} ,t ; \mathbf{r}^{\prime} ,t^{\prime}  \big)$. 
Note that we have numerically verified that Eq.~\eqref{eq:varphi_RF} (or equivalently Eq.~\eqref{eq:varphi_LF}) reproduces with great fidelity the ionization probabilities of~\cite{Barth2011} and~\cite{Barth2014}. Analytic expressions can be found in~\cite{Dubois2preparation}.
On short timescales, around the peak amplitude of the laser field, the pulse envelop is $f(t){=}1$, and the Green function in the RF becomes invariant under translation in time~\cite{SuppMat}
\begin{equation}
    \widetilde{G} (\mathbf{r},t ; \mathbf{r}^{\prime} ,t^{\prime}) = \widetilde{G} (\mathbf{r},t-t^{\prime} ; \mathbf{r}^{\prime} ,0),
\end{equation}
regardless of the interaction potential.
This clearly indicates that the energy of the ionizing wavepacket is conserved during tunnel ionization. The time-integral in~\eqref{eq:varphi_RF} can be substituted by the time-independent Green function $\widetilde{G}(\mathbf{r},\mathbf{r}^{\prime}; {-}\widetilde{I}_p)$ propagating the electron from $\mathbf{r}^{\prime}$ to $\mathbf{r}$ on a constant energy level ${-}\widetilde{I}_p{=}{-}(I_p{+}m\omega)$. The complex tajectories obtained by the saddle point approximation~\cite{Lewenstein1994} are depicted by dotted lines in Figs.~\ref{fig:LF_RF}b and~\ref{fig:LF_RF}d. They reproduce well the final energy in the LF and in the RF, and show that the energy of the electron in the RF does not change during tunneling.

As a final remark we would like to add, that the present scheme of adiabatic time-dependent motion in the RF can be extended to potentials which are not rotationally symmetric, as long as the resulting time-dependence of the potential in the rotating frame is slow, i.e. adiabatic. Generally speaking, this will be the case for potentials smoothly varying in space, which is the typical case for molecules and more complex systems.

\paragraph*{Summary}
We have shown that electrons subjected to strong CP laser pulses obey classical conservation laws, confirmed by semiclassical treatments.
These conservation laws offer a clear characterization of the tunnel ionization process, and provide a powerful tool for deeper analysis.
In addition, analysis in the rotating frame together with the conservation laws offer a promising avenue for predicting and controlling the phase-space configuration of the electron after tunnel ionization for more complex systems, such as molecules.

\paragraph*{Acknowledgements}
JD acknowledges ATTOCOM funded by the Agence Nationale de la Recherche.
JD acknowledges Kieran Fraser, Panos Giannakeas, Andrew Hunter and Gabriel Lando for fruitful discussions.

\end{document}